\documentclass[preprint]{aastex}
\slugcomment{Accepted for Publication in the Astronomical Journal}

\shorttitle{Scl-dE1 GC1: An Extended Globular Cluster}

\shortauthors{Da Costa et al}

\begin{document}

\title{Scl-dE1 GC1: An Extended Globular Cluster in a Low-Luminosity Dwarf 
Elliptical Galaxy\footnote{Based on observations made with the NASA/ESA Hubble 
Space Telescope, obtained at the 
Space Telescope Science Institute, which is operated by the Association of 
Universities for Research
in Astronomy, Inc., under NASA contract NAS 5-26555.  These observations are 
associated with
program GO-10503.}}

\author{G. S. Da Costa\altaffilmark{1}, E. K. Grebel\altaffilmark{2},  H. 
Jerjen\altaffilmark{1},  
M. Rejkuba\altaffilmark{3}, and M. E. Sharina\altaffilmark{4} }

\altaffiltext{1}{Research School of Astronomy \& Astrophysics, The Australian 
National 
University, Mt~Stromlo Observatory, via Cotter Rd, Weston, ACT 2611,
Australia}

\altaffiltext{2}{Astronomisches Rechen-Institut, Zentrum f\"{u}r Astronomie der 
Universit\"{a}t Heidelberg,
M\"{o}nchhofstr. 12-14, D-69120 Heidelberg, Germany}

\altaffiltext{3}{European Southern Observatory, Karl-Schwarzschild-Strasse 2, 
85748 Garching bei MŸnchen, Germany}

\altaffiltext{4}{Special Astrophysical Observatory, Nizhnii Arkhys 369167, 
Russia}

\begin{abstract}
We report the discovery from Hubble Space Telescope ACS images of an extended 
globular cluster, 
denoted by Scl-dE1 GC1, in the Sculptor Group dwarf Elliptical galaxy Scl-dE1 
(Sc22).  The
distance of the dE is determined as 4.3 $\pm$ 0.25 Mpc from the $I$ magnitude of 
the tip of the red 
giant branch in 
the color-magnitude diagram.  At this distance the half-light radius of Scl-dE1 
GC1 is $\sim$22pc, placing it among the largest clusters known, particularly for 
globular clusters associated with
dwarf galaxies.  The absolute magnitude of Scl-dE1 GC1 is M$_{V}$ = --6.7 and, 
to within
the photometric uncertainties of the data, the cluster stellar population 
appears indistinguishable 
from that of the dE.\@ We suggest that there may be two modes of globular 
cluster formation in
dwarf galaxies, a ``normal'' mode with half-light radii of typically 3 pc, and 
an ``extended'' mode
with half-light radii of $\sim$10 pc or more.
\end{abstract}


\section{Introduction}

Globular clusters  have generally been regarded as occupying distinctly 
different 
regions from dwarf galaxies in the multi-dimensional space formed from 
parameters such as central 
surface brightness, scale length, and total luminosity, mass or velocity 
dispersion.  Recent discoveries, 
however, have blurred the previous clear-cut separation.  For example, among the 
newly discovered
dwarf galaxy satellites of the Milky Way, there are systems such as Willman~1 
\citep{Wi05}, 
Segue~1 \citep{Be07} and Bo\"{o}tes~II \citep{WJ07} which have very low 
luminosities and small scale 
lengths that overlap with those of globular clusters \citep[e.g.][and the 
references
therein]{GG07,MJR08}.  Equally, there is increasing recognition of the existence 
of extended globular clusters with comparatively large scale lengths; the M31 
clusters discussed in 
\citet{HT05} and \citet{DM06}, and the M33 extended star cluster M33-EC1 
\citep{SV08}
 being prime examples.  Note that we argue that extended globular
clusters such as these are distinct from the diffuse stellar clusters discussed 
in \citet{PC06}, which have 
notably red (metal-rich) colors and which are spatially associated with galactic 
disks \citep{PC06}.
Similarly, we also regard extended globular clusters as distinct from the 
``faint fuzzies'', low luminosity
(M$_V$ $\gtrsim$ --7), large (half-light radii between 7 and 15 pc) star 
clusters that are spatially 
and kinematically associated with the disks of lenticular galaxies 
\citep[cf.][]{BL02}.

The origin of the extended globular clusters is not well understood.  
They are apparently comparatively rare objects, at least in luminous galaxies.  
For example,
\cite{GW07} indicate that in NGC~5128, extended clusters (defined as clusters 
with half-light radii, 
r$_h$, exceeding 8 pc) make up only $\sim$2.4\% of their sample.  Similarly, in 
the catalog of Milky 
Way globular cluster parameters \citep[][2003 Feb version]{WH96}, $\sim$9\% of 
clusters have 
r$_h$ exceeding 10 pc.  As has been known for some time \citep[e.g.][Fig. 
3]{vdbM04}, the majority of 
these extended clusters are found at large Galactocentric radii.  The results 
for four other spirals (M81, M83, NGC~6946, and M101)  show similar small 
fractions of extended clusters \citep{CWL04}, but 
intriguingly, in M51 $\sim$24\% (8 of 34) of the cluster candidates have r$_h$ 
$>$ 10 pc \citep{CWL04}. 
Extended globular clusters are also known to occur in dwarf galaxies,  for 
example, the Reticulum cluster in the LMC, Arp~2 in the Sagittarius dwarf, and 
cluster \#1 in Fornax all have r$_h$ exceeding 
10 pc \citep[e.g.][]{vdbM04}.

Clearly to understand the formation processes for these extended clusters, and 
particularly to investigate 
any dependence on environment or host galaxy type, structural parameters for 
larger samples of such
clusters are needed.  In this context, as already implicitly assumed, the 
appropriate structural parameter
for comparisons of clusters in different dynamical environments is the 
half-light, or effective, radius of a cluster, as it is generally regarded as 
being minimally affected by dynamical processes
\citep[e.g.][]{AH98}.   

In this paper we report the discovery on deep HST ACS images of an extended 
globular cluster 
in the dwarf Elliptical galaxy Scl-dE1 (Sc22), the first such cluster identified 
in a low luminosity dE\@.  
Scl-dE1 (Sc22) is a member of the Sculptor ``Group'', a loose
aggregation of galaxies that shows considerable extension along the 
line-of-sight \citep{JFB98, Ka03}. 
In the following section we describe the observations, while  $\S$ 3
discusses the color-magnitude diagram (CMD) of the dE and its distance from the 
$I$-mag of
the tip of the red giant branch.  The surface brightness profile of the globular 
cluster and 
associated parameters is also derived.  The results are discussed in $\S$ 4 in 
the context of globular clusters in galactic systems.

\section{Observations}

As part of GO program 10503, deep images of Scl-dE1 (Sc22) were obtained with 
the ACS 
instrument onboard HST in the $F606W$ (wide-$V$) and $F814W$ (wide-$I$) filters. 
 Each observation
consisted of a standard 4 point dither pattern with exposures of 1120 s (split 
into two to allow
cosmic ray rejection) at each pointing.  The images were processed through the 
standard ACS data
reduction pipeline and combined into single images for each filter using the 
MULTIDRIZZLE task.  
Subsequent analysis used these combined frames.  

\begin{figure}
\epsscale{0.5}
%\plottwo{MS_Scu22acs.ps}{MS_Scu22gc1.ps}
\plotone{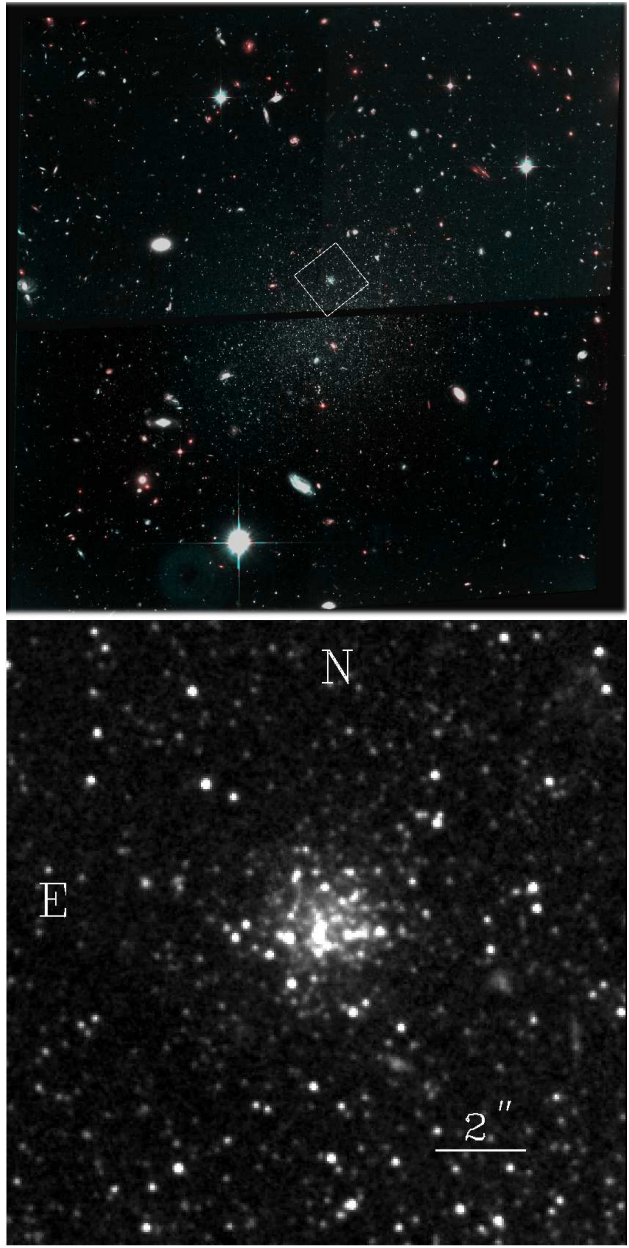}
\epsscale{0.5}
\plotone{newScu22gc_1.ps}
\caption{{\it Upper panel:} A color image of Scl-dE1(Sc22) from the $F606W$ 
and $F814W$ images.  The entire ACS field is shown.  North is $\sim$45$\arcdeg$ 
to the upper right and
East is $\sim$45$\arcdeg$ to the upper left.  The region outlined by the dotted 
line is reproduced
in the lower panel.  {\it Lower panel:} $F606W$ image of the globular cluster 
candidate Scl-dE1 
GC1.  North and East
are indicated and a 2$\arcsec$ scale bar is shown.  The cluster is 
clearly resolved into stars. \label{gcpic}}
\end{figure}

Based on relatively short exposure HST WFPC2 `snapshot' images, \citet{MS05} 
listed three candidate 
globular clusters in Scl-dE1 (Sc22).  Inspection of these candidates on our 
deeper ACS images,
however, reveals that all three candidates are background galaxies.  
Nevertheless, an additional globular 
cluster candidate is readily visible on the ACS frames\footnote{The candidate is 
visible on the WPC2
snapshot image, but was considered to be most likely a background cluster of 
galaxies, given the 
large apparent size.}, approximately 20$\arcsec$ NE of the center of the dE\@.  
We refer to this candidate 
as Scl-dE1 GC1.
A color image of the dwarf galaxy made from the $F606W$ and $F814W$ frames, with 
the cluster
candidate identified, is shown in the upper panel of Fig.\ \ref{gcpic}.  The 
lower panel shows an
enlargement of the vicinity of the cluster, which is clearly resolved into 
stars.
  
\section{Analysis}

\subsection{Color-magnitude diagrams}

The first step in the analysis was to determine a CMD for the stellar population 
of Scl-dE1 (Sc22).  
The stand-alone version of Stetson's DAOPHOT/ALLSTAR package \citep{PB87, PB94} 
was employed 
to carry out the photometric reductions on a pair of $F606W$ and $F814W$ frames. 
 For each frame 
typically 20-30 of the brighter stars were used 
to generate the PSF\@.  After the first ALLSTAR run, the subtracted frame was 
searched for stars missed
in the first FIND pass, which were then added into the input list and ALLSTAR 
run again.  This process was
repeated for a third time to ensure the maximum number of stars were measured.  
Aperture photometry
was then carried out on the PSF stars to determine the correction from the PSF 
magnitudes to aperture
magnitudes inside a 10pix (0.5$\arcsec$) radius aperture.  The calibration 
procedures outlined in 
\citet{S05} were then used to convert the photometry to the ACS VEGAMAG system.  
Objects with
discrepant CHI or SHARP parameters, or magnitude errors, relative to mean values 
 for their measured
magnitudes, were deleted from the photometry lists. The cleaned lists for the 
$F606W$ and $F814W$ 
images were then matched and the photometry converted to Johnson-Cousins $V$ and 
$I$ magnitudes
using the transformations in \citet{S05}.  For a typical red giant at $I$ 
$\approx$ 25 and $(V-I)$ $\approx$ 1.3, the photometric uncertainties are
approximately 0.04 and 0.06 mag for $I$ and $(V-I)$.  The systematic 
uncertainties that affect the overall 
zero points of the photometry are, for $I$ and $(V-I)$ respectively,  
0.015 and 0.03 mag from the uncertainty in the aperture corrections, and 
0.02 and 0.025 mag from the uncertainties in the transformations
given by \citet{S05}.

\begin{figure}
\epsscale{0.5}
\plotone{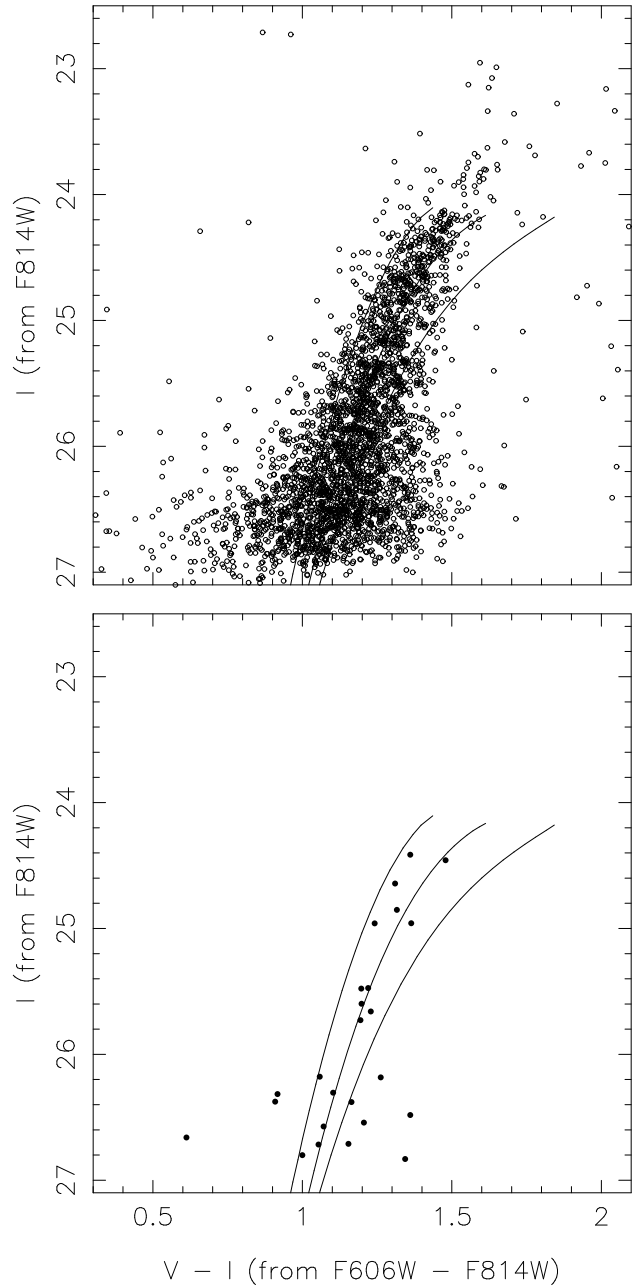}
\caption{(a) {\it Upper panel}. A color-magnitude for all stars on the HST/ACS 
images of Scl-dE1 (Sc22).
Shown also are the red giant branches for the Galactic globular clusters stars 
M15 ([Fe/H] = --2.17),
M2 (--1.58) and NGC~1851 (--1.16) from \citet{DA90} fitted assuming (m--M)$_{I}$ 
= 28.20 and 
E($V-I$) = 0.02 mag.  (b) {\it Lower panel}.  Color-magnitude diagram for stars 
lying within a radius
of 50 pix (2.5\arcsec) from the center of Scl-dE1 GC1.  The giant branches are 
as for the upper panel.
 \label{cmd_fig}}
\end{figure}

The resulting CMD for the dE is shown in the upper panel of Fig.\ \ref{cmd_fig}. 
  Detailed discussion 
of the stellar population
of Scl-dE1 (Sc22) will be included in a subsequent paper.  Here we note only 
that the stellar population
appears to be dominated by old, metal-poor stars, as expected given the 
morphological classification of
this dwarf and its apparent lack of neutral hydrogen \citep[cf.][]{AB05}.  An 
$I$-band luminosity 
function generated
from the data of the upper panel of Fig.\ \ref{cmd_fig} reveals that the tip of 
the red giant branch
(TRGB) is at $I$ = 24.15 $\pm$ 0.12,
where the error includes contributions from the uncertainty in locating the 
actual tip given the
number of stars involved and the photometric errors, as well from the zero point 
and aperture correction 
uncertainties.   \citet[][see also \citet{TS08}]{LR07} discuss the calibration 
of the absolute magnitude of
the TRGB and give M$_{I}^{TRGB}$ = --4.05 $\pm$ 0.02 at $(V-I)_{0}$ = 1.6, 
corresponding to 
[Fe/H] $\approx$ --1.5 for an old stellar population.  This value is remarkably 
close to the original
calibration of \citet{DA90} which gives M$_{I}^{TRGB}$ = --4.02 for the same 
parameters.  Adopting
M$_{I}^{TRGB}$ = --4.05, then gives (m--M)$_{I}$ = 28.20 which, for a reddening 
E(B--V) = 0.015
\citep{Sc98} and A$_{I}$=1.86~E(B--V), yields a distance for Scl-dE1 (Sc22) of 
4.3 $\pm$ 0.25 Mpc.
This value is consistent with that, 4.2 $\pm$ 0.45 Mpc, given by \citet{Ka03} 
from the $I$-mag of the
TRGB in a CMD based on HST snapshot images.  
%Both this value and that derived here considerably
%exceed the distance, 2.7 $\pm$ 0.2 Mpc, given by \citet{JFB98} from a surface 
%brightness fluctuation
%analysis.  

With the distance determined, we can estimate the mean metallicity of Scl-dE1 
(Sc22) by
comparing the mean color of the red giant branch with those of Galactic globular 
clusters.  Using the
calibration given in \cite{NC98} for $(V-I)_{0,-3.5}$, the dereddened color of 
the giant branch at M$_{I}$ = --3.5, the mean metal abundance of Scl-dE1 (Sc22) 
is $\langle$[Fe/H]$\rangle$ = --1.73 $\pm$  0.17,
where the uncertainty includes the statistical error in the mean color, the 
aperture correction and photometric transformation
uncertainties, the uncertainty in the distance and the uncertainty in the 
calibration.  This mean abundance is 
again consistent with that, $\langle$[Fe/H]$\rangle$ = --1.5 $\pm$ 0.3, given by 
\citet{Ka03}.  

In the lower panel of Fig.\ \ref{cmd_fig} we show the photometry for stars 
selected to lie within a radius
of 50 pix (2.5$\arcsec$) from the center of the candidate globular cluster.  The 
relatively small number of  
stars in this CMD preclude any definite conclusions regarding the stellar 
population of the cluster,
but it is clearly not significantly different from that of the field population 
of the dE, i.e.\ it is also old and 
metal-poor, supporting the interpretation of the cluster as a definite globular 
cluster.  A first order estimate 
of the mean abundance of Scl-dE1 GC1 can be obtained by interpolating 
the $(V-I)$ colors of the stars brighter than $I$ = 26 in the lower panel of 
Fig.\ \ref{cmd_fig} within the 
frame of the Galactic globular cluster giant branches depicted.  This yields 
$\langle$[Fe/H]$_{GC1}\rangle$ $\approx$ --1.7 $\pm$ 0.3 dex, again not notably 
different from the
dE population as a whole.  This lack of any clear metallicity difference between 
the cluster and field stars
is similar to the situation in the dwarf irregular galaxy WLM where the 
metallicity of the single luminous 
globular cluster is similar to that of the old RGB stars \citep[e.g.][]{Do02}.

\subsection{Surface photometry of Scl-dE1 GC1}

A surface brightness profile of the globular cluster was generated by carrying 
out aperture photometry
with a series of apertures of increasing size centered on the cluster on the 
$F606W$ and $F814W$
frames.  The initial estimate of the cluster center was determined by eye and 
then refined via application
of the centroiding technique in the IRAF {\it apphot} routine.  The uncertainty 
in the cluster center
position is of order $\pm$2 pix 
(0.1$\arcsec$) in each co-ordinate.  The background was determined from the 
signal in an
annulus centered on the cluster, with inner and outer radii of 60 and 120 pixels 
(3--6$\arcsec$).
A plot of the (background subtracted) concentric aperture magnitudes against 
aperture radius
then readily leads to an estimate of the total magnitude for the cluster.  After 
conversion 
to the $V, I$-bands, we find $V$ = 21.55 $\pm$ 0.05 and $V-I$ = 1.02 $\pm$ 0.02 
for Scl-dE1 GC1,
where the error includes the uncertainty in the background determination.  
For the distance modulus and reddening given above, this corresponds to M$_{V}$ 
= --6.67 $\pm$ 0.13,
where the dominant error is the uncertainty in the distance modulus.  Scl-dE1 
GC1 is thus somewhat
fainter than the peak of the globular cluster luminosity function in dE systems 
(M$_{V}$ = --7.3, \citet{ML07}), 
but not substantially so given that the 1-$\sigma$ width of the function is 
$\sim$1 mag \citep{ML07}.
The $(V-I)_{0}$ color of 1.00 $\pm$ 0.02 (measurement error only) is somewhat 
redder than might be
expected for the mean metallicity estimate derived above: for example, the 
relation given in \citet{BH00} 
would predict $(V-I)_{0}$ = 0.89 $\pm$ 0.03 for $\langle$[Fe/H]$_{GC1}\rangle$ 
$\approx$ --1.7 $\pm$ 0.2 
dex.  However, given that the relatively low luminosity of the cluster results 
in a sizable statistical
uncertainty in the measured color, and given the uncertainty in the abundance 
estimate, this difference 
does not seem particularly significant.

Once the integrated magnitude profile and its asymptote are established, it is 
straightforward to fix
the value of r$_{h}$, the radius containing half the total light.  For the 
$V$-band data, we find r$_{h}$
= 1.06$\arcsec$ $\pm$ 0.05, or 22.0 $\pm$ 1.5 pc for Scl-dE1 GC1, where the 
latter uncertainty includes 
the distance uncertainty.  {\it This is an exceptionally large value}: for 
example, in the \citet{WH96} catalog 
(2003 Feb version) for Milky Way globular clusters, only Pal~14 with r$_h$=24.6 
pc exceeds this value,
with the next three largest clusters being Pal~5 (r$_h$=19.9 pc), NGC~2419 (17.8 
pc) and Pal~3 (17.7 pc). 
We shall return to this point in the following section.

\begin{figure}
\epsscale{0.7}
\includegraphics[angle=-90]{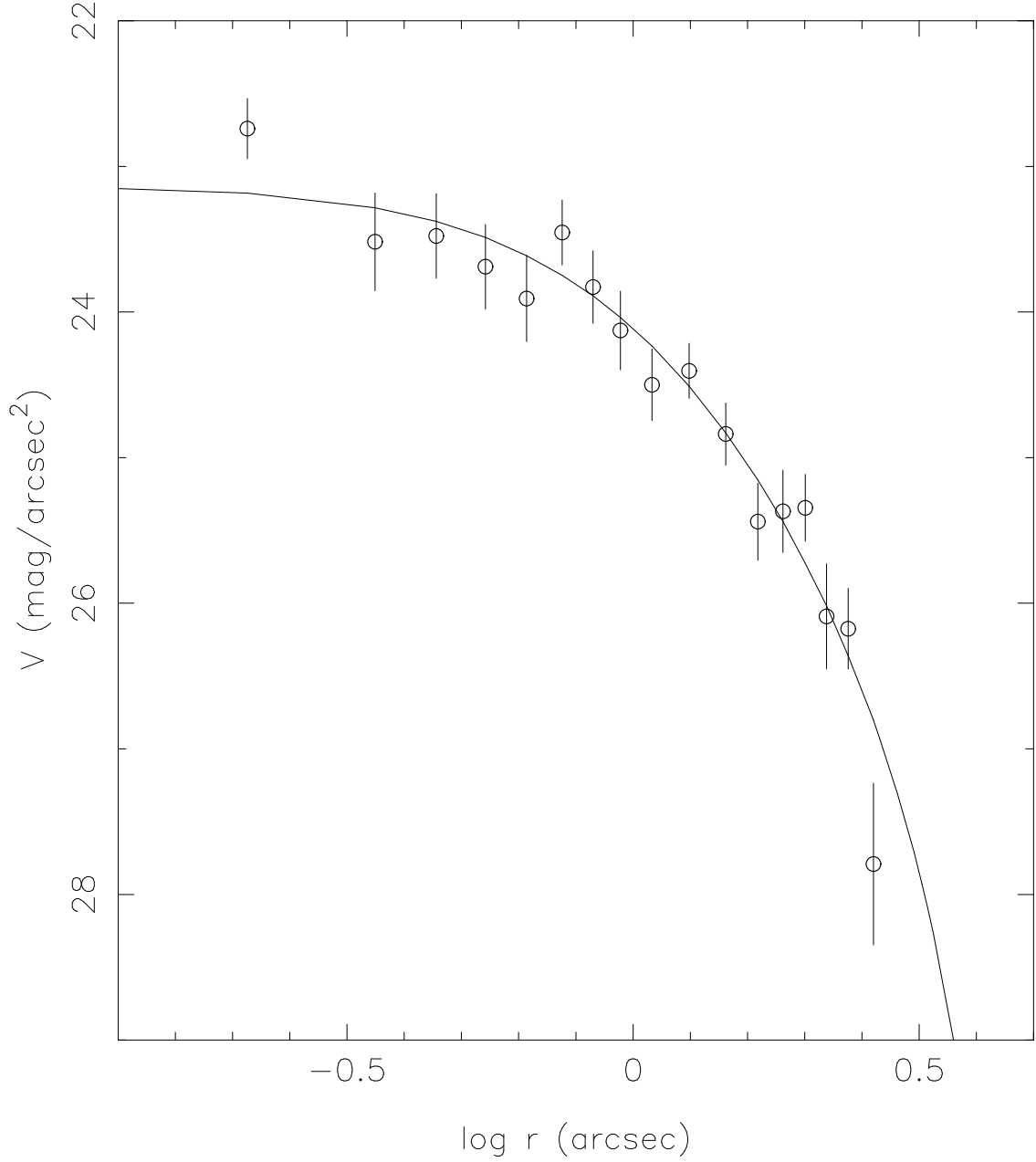}
\caption{ The $V$-band surface brightness profile of Scl-dE1 GC1 derived from 
concentric aperture
photometry.  Shown also is a \citet{IK62} profile with a core radius of 
1.0$\arcsec$ (21.2 pc), a 
concentration index $c$=0.65, and central surface brightness of 23.1 
$V$mag/arcsec$^2$. \label{sbprof_fig}}
\end{figure}

The concentric aperture photometry can be readily turned into a surface 
brightness profile using
standard techniques, e.g.\ \citet{GDC79}.  The resulting profile is shown in 
Fig.\ \ref{sbprof_fig}, where  
the error bars are the sampling errors, i.e.\ the uncertainty resulting from the 
finite number of stars 
contributing to the light in each annulus.  They are calculated following 
\citet{II76}.  Shown
also in the figure is a fit of a \citet{IK62} analytic profile with parameters 
central surface brightness $\mu_0$
= 23.1 $V$mag/arcsec$^2$, core radius r$_c$ = 1.0$\arcsec$ (21.2 pc), and a 
concentration index $c$ 
= log(r$_t$/r$_c$)=0.65.  Given the uncertainty in the individual points, this 
profile is an adequate 
representation of the data.  The parameters for Scl-dE1 GC1 are summarized in 
Table \ref{table1}.

\begin{deluxetable}{ll}
\tablewidth{0pt}
%\tablecaption{Parameters of Scl-dE1 GC1 \label{table1}}
\tablecolumns{2}
\tablehead{Parameters of Scl-dE1 GC1 \label{table1}}
\startdata
Position $\alpha, \delta$ (J2000) & 00 23 52.69, --24 41 58.0\\
Distance from galaxy center & $\sim$20$\arcsec$ NE ($\sim$415 pc)\\
(m--M)$_0$ & 28.17 $\pm$ 0.12 (4.3 $\pm$ 0.25 Mpc)\\
M$_V$ & --6.67 $\pm$ 0.13\\
$(V-I)_0$ & 1.00 $\pm$ 0.02\\
$\mu_0$ ($V$mag/arcsec$^2$) & 23.1 $\pm$ 0.4\\
c & 0.65 $\pm$ 0.1\\
r$_c$ & 1.0$\arcsec \pm 0.1$ ($21.2 \pm 2.1$ pc)\\
r$_h$ & 1.06$\arcsec \pm 0.05$ ($22.0 \pm 1.5$ pc)\\
$[$Fe/H$]$ & --1.7 $\pm$ 0.3 (from RGB) \\

\enddata
\end{deluxetable}

\section{Discussion}

The {\it specific frequency S$_N$}, i.e.\ the number of globular clusters in a 
galaxy per unit luminosity,
is known to increase with decreasing luminosity in both nucleated and 
non-nucleated dE
galaxies, and to be higher in nucleated systems \citep[e.g.][]{ML07}.  With its 
one globular cluster
and an absolute visual magnitude M$_{V}$ $\approx$ --11.5, the non-nucleated dE
Scl-dE1 (Sc22) has S$_N$ $\approx$ 25.
This is a large but not exceptional value: \citet{ML07} list two Virgo cluster 
non-nucleated dEs with 
M$_V$ $\approx$ --13 and S$_N$ $\approx$ 20, while the Fornax dwarf Spheroidal 
galaxy, with its 
five globular clusters, has S$_N$ $\approx$ 26.  Nevertheless, Scl-dE1 (Sc22) is 
one of the lowest 
luminosity dEs known to possess a globular cluster -- for example, in 
\citet{G08b}, the lowest
luminosity dEs with globular cluster candidates listed are the M81 group object 
IKN with M$_V$ 
$\approx$ --11.5, and the Cen~A group dSph [KK2000] 55 (KKS 55), which has 
M$_V$ $\approx$ --11.2.

\begin{figure}
%\epsscale{0.1}
\includegraphics[angle=-90,scale=0.7]{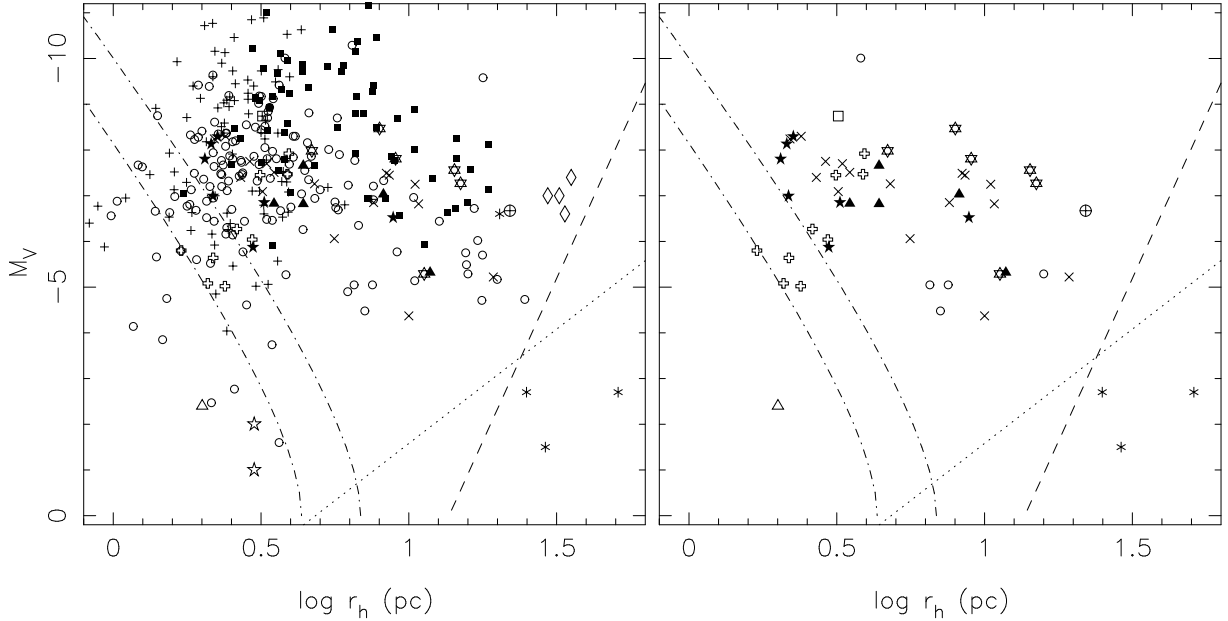}
\caption{Absolute magnitudes for stellar systems as a function of the logarithm 
of the half-light
radius in parsecs.  Scl-dE1 GC1 is the circled plus sign, the WLM cluster is the 
open square, the
open diamonds are the extended M31 clusters from \citet{DM06}, the open triangle 
is Whiting~1
and the open 5-pt stars are the Koposov clusters.  LMC clusters are x-signs, SMC 
clusters are 6-pt
stars, and filled triangles are the Fornax clusters.  The filled 5-pt stars are 
NGC~205 clusters and 
NGC~147 and NGC~185 clusters are the open plus symbols.  M33-EC1 is the asterisk 
symbol
adjacent to Scl-dE1. Milky Way globular clusters
are open circles, M31 clusters are plus signs and NGC~5128 clusters are filled 
squares.  The asterisks
to the right of the `Shapley line' \citep[cf.][]{vdb08}, shown by the dashed
line, are, in order of increasing log r$_{h}$, the dwarf galaxies Willman~1, 
Segue~1 and Bootes~II.  The dotted line represents a constant effective surface 
brightness of 27.0 $V$ mag per arcsec$^2$.  The dot-dash lines show where 20 
(upper line)
and 40 (lower line) times the 2-body relaxation time at the half-mass radius is 
equal to 12 Gyr.
The right panel shows the same data as the left panel, but is restricted to 
globular clusters in
dwarf galaxies.  Here open circles are used for the clusters associated with the 
Sgr dwarf, otherwise
the symbols are the same as the left panel.
\label{rh_fig}}
\end{figure}

Turning now to the large size of Scl-dE1 GC1, in the left panel of Fig.\ 
\ref{rh_fig} we compare 
our results to (log r$_h$, M$_V$)
values for globular clusters in a number of different galactic systems.   The 
sources of the data are:
Milky Way globular clusters, \citep[][2003 Feb version]{WH96}; Koposov clusters 
and Whiting~1, 
\citep{KdJ07}; LMC clusters \citep{vdbM04}; SMC clusters \citep{KG08b}; Fornax 
dSph clusters, 
\citep{vdbM04}; M33-EC1, \citep{SV08}; M31 clusters, \citep{BMc07}; M31 extended 
clusters, 
\citep{DM06}; NGC~205 clusters, 
\citep{BMc07}; NGC~147 and NGC~185 clusters, \citep{SP08}; WLM globular cluster 
\citep{SCC06}; 
and NGC~5128 clusters, \citep[][Appendix]{GG06,McB08}.  With the exception of 
Whiting 1, which
has an age of $\sim$6.5 Gyr \citep{CZ07}, and the SMC clusters Lindsay~1, 
Kron~3, NGC~339,
NGC~416 and Lindsay~38, which have ages of 7.5, 6.5, 6, 6 and 6.5 Gyr, 
respectively, \citep{KG08a},
all the clusters plotted are ``old'' objects, i.e.\ ages exceeding 10 Gyr.
 
Shown also on the plot are (log r$_h$, M$_V$) values for three recently 
discovered very low-luminosity
dwarf galaxies, Willman~1, Segue~1 and Bootes~II, with the data taken from 
\citet{MJR08}.  These dwarf
galaxies have r$_h$ values comparable with those of the largest globular 
clusters, but the clusters are
more luminous.  More luminous dwarf galaxies than those plotted have larger  
r$_h$ values and lie 
to the right of the ``Shapley line'' from \citet{vdb08}, which 
is shown as the dashed line and which (empirically) appears to separate globular 
clusters from dwarf galaxies.  A constant effective surface brightness of 
$\sim$27.0 $V$ mag per arcsec$^2$ is also shown as the dotted line.  This line 
also separates the dwarf galaxies and globular 
clusters, at least for the dwarf galaxies with r$_h$ values less than $\sim$100 
pc.  
The right panel repeats the data but is restricted solely to globular clusters 
in dwarf systems (LMC, SMC, Sgr, Fornax, NGC~147, 185, 205 and WLM in addition 
to Scl-dE1).  

It is immediately apparent from these plots that Scl-dE1 GC1 is an unusual 
object: there are only a small
number of globular clusters of comparable size and only the extended clusters 
found in M31 
\citep{DM06} definitely exceed it in size.  Nevertheless, Scl-dE1 GC1 remains a 
factor of $\sim$10 
smaller than dwarf galaxy companions to the Milky Way such as Boo~I, Her and 
UMa~I, which have
comparable luminosities to the cluster \citep{MJR08}.  The separation between 
star clusters and 
dwarf galaxies at fixed size (clusters are brighter) or fixed luminosity 
(clusters are smaller) remains
in place.

Are there any clues to the origin of Scl-dE1 GC1?  Stellar systems are subject 
to a number of
dynamical processes that can lead to their disruption.  For example, 2-body
relaxation processes can lead to disruption of a cluster on a timescale of 
$\sim$30--50$t_{rh}$, where 
$t_{rh}$ is the 2-body relaxation time computed at the half-mass radius 
\citep[e.g.][]{GO97}.  
We show as dot-dash lines in Fig.\ \ref{rh_fig} the relation between log r$_h$ 
and M$_V$ for the case where 20$t_{rh}$ (upper line) and 40$t_{rh}$ (lower line) 
are equated with 
a Hubble time of 12 Gyr.  Clusters lying below these `survival lines'  are 
strongly affected by this
process.  In calculating these lines we have assumed a constant mass-to-light 
ratio of 1.6 in solar
units, an average stellar mass of 0.35M$_{\sun}$ and a coulomb factor {\it 
ln}($\Lambda$) equal to
{\it ln}(0.02~N)\citep[cf.][]{GO97, G08b}.   This disruption process is clearly 
unimportant for
Scl-dE1 GC1.  

For Galactic globular clusters, the tidal gravitational shocks that occur when 
the 
cluster passes through the disk of the Galaxy, or passes close to the Galactic 
bulge, can contribute
at least as much as 2-body relaxation to cluster disruption 
\citep[e.g.][]{GL99}.  The disruption
timescale from gravitational shocks is proportional to the mass of the cluster 
divided by $r_{h}^{3}$
\citep[e.g.][]{GL99}, so that large low-mass clusters are particularly at risk 
from this disruption
mechanism.  This will be especially the case for clusters with small 
Galactocentric radii. 
Indeed, while the fact that the extended globular clusters in the Milky Way halo 
are predominantly
at large Galactocentric radii is sometimes used to argue that these clusters are
accreted objects \citep[e.g.][]{MG04,vdbM04}, it is actually the low density 
environment at these large 
distances, where they are not subject to substantial tidal shock effects, that 
has ensured their survival.     
It is not surprising then that the M31 extended clusters are also outer objects 
with projected 
galactocentric distances $15 \lesssim R_{p} \lesssim 60$ kpc \citep{DM06} and 
that  
the most extended LMC cluster (Reticulum) lies in the extreme outskirts of the 
LMC \citep[cf.][]{vdbM04}.  The extended M33 cluster M33-EC1 also lies well
beyond the disk of M33, at a projected distance of 12.5 kpc from the center
of the galaxy \citep{SV08}.
Moreover, in the NGC~5128 samples used here (see above), globular clusters with 
r$_h$ exceeding 
10~pc are found at all projected radii from the center of the galaxy beyond 5 
kpc, but are apparently 
notably more frequent beyond R$_{p}$ = 25 kpc \citep[cf.\ Fig.\ 9 of][]{GG06}.   
Thus a relatively benign
gravitational environment would seem to be a prerequisite for the occurrence of 
extended star clusters,
and that is the case for Scl-dE1 GC1. 
Regardless of whether Scl-dE1 (Sc22) contains a dark halo or not, the density 
profile of
this (non-nucleated) dwarf is smoothly varying \citep[e.g.][]{JBF00}, and it is 
presumably this 
consequent lack of substantial tidal effects that allowed such a large cluster 
as Scl-dE1 GC1 to form 
and survive.  

\begin{figure}
%\epsscale{0.3}
\includegraphics[angle=-90,scale=0.65]{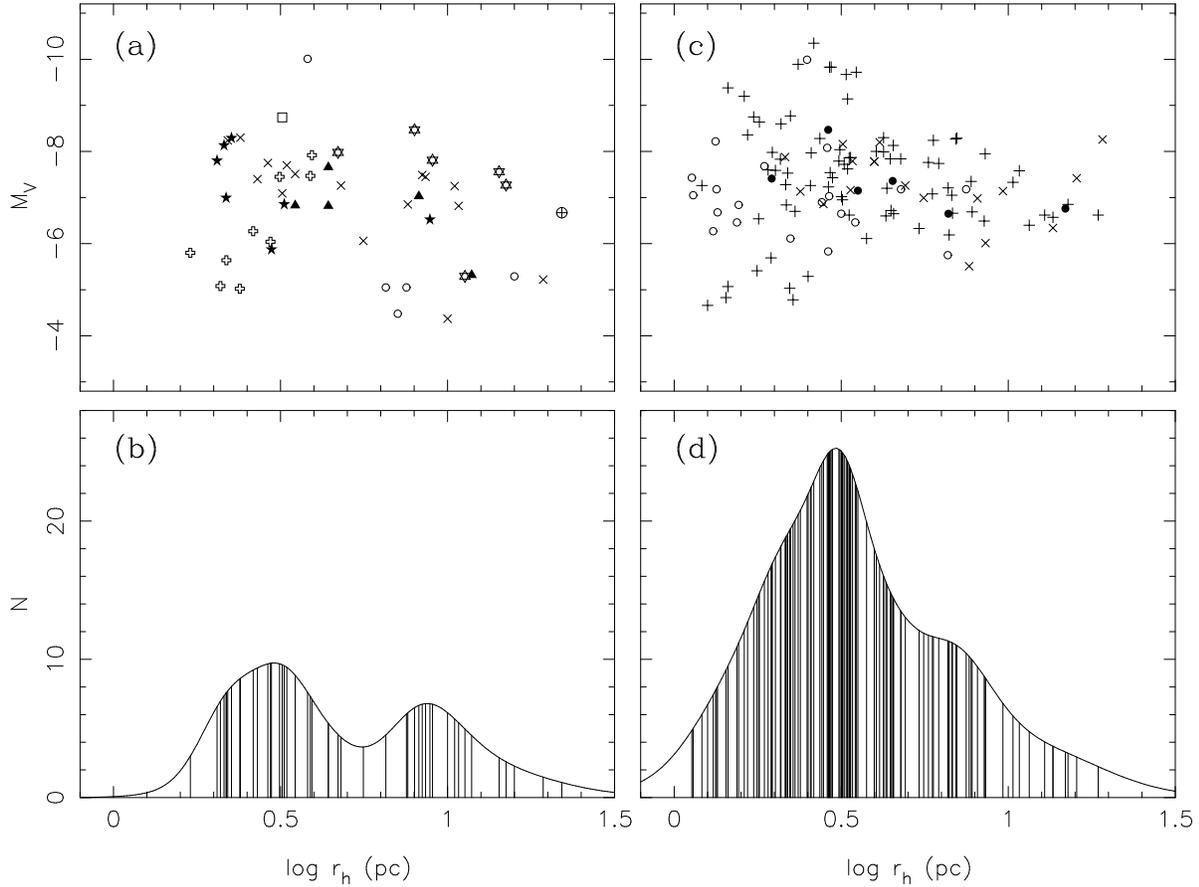}
\caption{(a) upper left panel: globular cluster absolute magnitudes as a 
function of the logarithm of the half-light radius in parsecs from the right 
panel of Fig.\ \ref{rh_fig}, sans Whiting 1.  Symbol definition 
unaltered from that figure.  (b) lower left panel: the distribution of log 
r$_h$ values for this sample 
resulting from the application of an adaptive kernel estimator.  The 
distribution is clearly bimodal.  
(c) upper right panel: globular cluster data selected from \cite{G08b} as 
discussed in the text.  
Plus symbols, filled circles, open circles and x-signs are for globular clusters 
in dIrr, dSph, dE and 
Sm galaxies, respectively.
(d) lower right panel: the corresponding log r$_h$ distribution.  This 
distribution also shows signs
of a bimodality.  
\label{rh_dist_fig}}
\end{figure}

Further, although there are undoubtedly significant selection effects governing 
the distribution of
points in the right panel of Fig.\ \ref{rh_fig}, there is a hint that the 
distribution of log r$_h$ values
for the globular clusters in these dwarf galaxies, which are predominantly Local 
Group systems, 
is bimodal.  There are apparent peaks in the distribution at $\sim$0.48 (3 pc) 
and $\sim$1.0 (10 pc), and a dearth of clusters at log r$_h$ $\approx$ 0.75 (5.5 
pc).   In the upper
left panel of Fig.\ \ref{rh_dist_fig} we repeat the data from the right panel of 
Fig.\ \ref{rh_fig} (sans Whiting 1) and in the lower panel left show the 
distribution of log r$_h$ values 
for these 50 clusters that results from application of the adaptive kernel 
estimator discussed in 
\cite{VF94}.  The distribution is indeed bimodal: there are peaks in the 
distribution at log r$_h$
$\approx$ 0.48 and log r$_h$ $\approx$ 0.94, and if log r$_h$ = 0.75 is used as 
the dividing
line, there is a $\sim$1.5 to 1 split between the two populations.
We interpret this as indicating there may be two modes of cluster formation, 
with the less common 
`extended cluster'  mode being relatively more frequent in the gravitationally 
smoother environment 
of dwarf galaxies compared to larger galaxies.  Such a scenario is consistent 
with the cluster formation
processes discussed in \citet{BE08}.
 
 The recent results of \citet{G08b} provide an independent means to investigate 
our suggestion.
 These authors have used archival HST/ACS images to investigate the globular 
cluster
 populations of 68 dwarf galaxies with distances less than 12 Mpc.  
 No Local Group dwarfs (nor Scl-dE1) are included in their  sample.  We show in 
the upper right panel 
 of Fig.\ \ref{rh_dist_fig}, the location of 126 globular cluster
 candidates in the (log r$h$, M$_V$) plane
 from the \citet{G08b} data set, noting that the vast majority of these clusters
 are not resolved into stars.
 The clusters were chosen to have 0.7 $\leq$ $(V-I)_{0}$ $\leq$ 1.1 and M$_V$ 
$\geq$ --11 to select
 old globular clusters of similar characteristics to those of the Local Group 
(plus Scl-dE1) dwarf 
 systems.  This sample comes
 from the globular cluster systems of 24 dIrrs, 2 dSphs, 2 dEs and 4 Sm 
galaxies.  Clearly it
 includes a number of extended clusters, such as cluster 3 in the M81 group dSph 
IKN, which
 has r$_h$ = 14.8 $\pm$ 0.8 pc, cluster 2 in the Cen~A/M83 group dSph/I 
NGC~5237, which has
 r$_h$ = 15.1 $\pm$ 0.2 pc, and cluster 10 in the field Sm NGC~4605, which has 
 r$_h$ = 19.2 $\pm$ 0.3 pc \citep{G08b}.
 The lower right panel of Fig.\ \ref{rh_dist_fig} shows the inferred log r$_h$ 
distribution for these
 126 clusters, again using an adaptive kernel estimator.  This distribution 
shows a peak at log r$_h$ 
 $\approx$ 0.48 (3 pc), essentially the same value as the lower log r$_h$ peak 
for the Local Group
 sample.  There is also a less prominent peak at log r$_h$ $\approx$ 0.84, which 
is at a
 somewhat lower value than the location of the second peak in the distribution 
in the left panel
 of the figure.  If we split our \citet{G08b} sample at log r$_h$ = 0.75 as 
above, the numbers of 
 clusters below and above this limit are in the ratio 2.9 to 1, which is 
substantially more than for
 the Local Group (plus Scl-dE1) sample.  This difference may be due in part to 
selection effects as
 \citet{G08b} note the more extended clusters suffer from stronger 
incompleteness.  Nevertheless, the distribution in the lower right panel of 
Fig.\ \ref{rh_dist_fig} is not unimodal: a K-S
 test indicates that there is a less than 10\% probability that the sample of 
\cite{G08b}
 clusters shown is drawn from a gaussian in log r$_h$ with mean 0.5 and $\sigma$ 
= 0.25 dex, values
 which provide an adequate representation of the distribution of log r$_h$ 
values equal to and
 less than the peak value.   Thus our 
 suggestion of the possibility of two modes of cluster formation in dwarf 
galaxies is by no means
 ruled out by the \citet{G08b} data set.   
 
 Similarly, while {\it explicitly excluding potential extended
 globular clusters} by eliminating cluster candidates with r$_{h}$ $>$ $\sim$10 
pc, the
 study of globular cluster sizes in Virgo early-type galaxies of \citet{JC05}
 nevertheless finds that the (surface brightness corrected) mean half-light 
radius for globular
 cluster systems increases with decreasing galaxy luminosity.  Inspection of 
their Fig.\ 10
 suggests this effect is not due to a shift in the main peak of the distribution 
at r$_h$ $\approx$ 2.7 pc, 
 but rather results from an increased fraction of larger clusters for less 
luminous (bluer) galaxies
 \citep{JC05}.
 
We note that the proposed bimodal cluster size distribution would not be 
expected to be as evident 
in the size distribution of globular clusters in luminous galaxies (cf.\ the 
left panel of Fig.\ \ref{rh_fig}
and the discussion in \S1)  because tidal and other dynamical effects act more 
strongly to modify 
the distribution of globular cluster
sizes in larger galaxies.  Consequently, the extended clusters seen in large 
galaxies may well indeed
be accreted objects, having formed in (now disrupted) dwarf systems.   In this 
scenario Scl-dE1 (Sc22) 
and the dIrr WLM make an interesting comparison: both have a single globular 
cluster, but from the different proposed modes: in Scl-dE1 (Sc22) the cluster is 
extended while in WLM it is relatively compact.    

\acknowledgments
We are grateful to our colleague Bruno Binggeli for his comments on the original 
manuscript. 
EKG acknowledges support from SNF grants 200020-113697 and 200020-122140 while 
MS acknowledges support from 
RFBR grant 08-02-00627.

{\it Facilities:} \facility{HST}

%\pagebreak

\end{document}